\documentstyle[12pt]{article}
\begin{document}

\begin{titlepage}

\begin{center}
{\bf Budker Institute of Nuclear Physics}
\end{center}

\vspace{1cm}

\begin{flushright}
BINP 95-13\\
February 1995
\end{flushright}
\bigskip
\begin{center}
{\bf WHAT DO WE LEARN FROM ATOMIC PHYSICS\\
ABOUT FUNDAMENTAL SYMMETRIES IN NUCLEI AND PARTICLES\\
Talk at Moriond 95 Workshop:\\
Dark Matter in Cosmology, Clocks and\\
Tests of Fundamental Laws}
\end{center}

\begin{center}
I.B. Khriplovich\footnote{e-mail address: khriplovich@inp.nsk.su}
\end{center}
\begin{center}
Budker Institute of Nuclear Physics, 630090 Novosibirsk,
Russia
\end{center}

\bigskip

\begin{abstract}
Atomic experiments bring meaningful and valuable information on
fundamental symmetries. The hypothesis of a large ($\sim 100$ eV)
P-odd weak matrix element between single-particle states in heavy
nuclei is inconsistent with the results of atomic PNC experiments.
Upper limits on CP-violation obtained in atomic and molecular
spectroscopy are as informative as those established in neutron
physics. Very strict upper limits on T-odd, P-even interactions
(nucleon-nucleon, electron-nucleon, electron-electron, and
$\beta$-decay) are derived from the same atomic and neutron
experiments.
\end{abstract}

\vspace{3cm}

\end{titlepage}

1. The scattering cross-sections of longitudinally polarized epithermal
(1 - 1000 eV) neutrons from heavy nuclei at $p_{1/2}$ resonances have
large longitudinal asymmetry. For a long time the most natural
explanation of the effect was based on the statistical model of the
compound nuclei. In fact, not only the explanation, but the very
prediction of the huge magnitude of this asymmetry (together with the
nuclei most suitable for the experiments) was made theoretically on
the basis of this model \cite{sf}.

An obvious prediction of the statistical model is that after
averaging over resonances, the asymmetry should vanish. However, few
years ago it was discovered \cite{fb} that all seven asymmetries
measured in $^{232}$Th have the same, positive sign.

All the attempts [3 - 7] to explain a common sign require the
magnitude of the weak interaction matrix element, mixing
opposite-parity nuclear levels, to be extremely large,\newline $\sim
100$ eV.
\footnote{The only exception known to me is recent paper \cite{fz}
where large octupole deformation of nucleus is discussed as a possible
explanation of this regularity.}
The same assumption seems to be necessary to explain unexpectedly
large P-odd correlations observed in M\"{o}ssbauer transitions in
$^{119}$Sn and $^{57}$Fe \cite{ts,ts1}.

In Ref. \cite{sb} it was pointed out that such a large magnitude of
the weak mixing can be checked in an independent experiment. The
proposal is to measure PNC asymmetry in the M4 $\gamma$-transition
between the (predominantly) single-particle states $\;1i\; 13/2^+\;$ and
$\;2f\; 5/2^-\;$ in $^{207}$Pb. The sensitivity of this experiment to the
weak matrix element value is expected to reach $5 - 13$ eV.

However, it was demonstrated recently \cite{dkt0} that close upper
limit on the weak mixing in $^{207}$Pb can be extracted already now
from the measurements of the PNC optical activity of atomic lead
vapour. The following upper limit was established at the 95\%
confidence level for the ratio of the nuclear-spin-dependent (NSD)
part of the optical activity to the main, nuclear-spin-independent
one \cite{mvm}:
\begin{equation}\label{ra}
\frac{P_{NSD}}{P} < 0.02
\end{equation}

In heavy atoms the NSD P-odd effects were shown to be induced mainly
by contact electromagnetic interaction of electrons with the anapole
moment of a nucleus which is its P-odd electromagnetic characteristic
induced by PNC nuclear forces \cite{fk,fks}. The result (\ref{ra})
leads to the following bound on the dimensionless anapole constant:
$\kappa(^{207}Pb) < 1$, and on the effective neutron PNC constant:
$g_n < 10$. The last constant is introduced via the effective P-odd
potential for an external nucleon:
\begin{equation}\label{we}
W=\frac{G}{\sqrt{2}}\;\frac{g}{2m}\;\vec{\sigma}[\vec{p}\rho(r)
+\rho(r)\vec{p}\;].
\end{equation}
Here $G=1.027\cdot 10^{-5} m^{-2}$ is the Fermi weak interaction
constant, $m$ is the proton mass, $\vec{\sigma}$ and $\vec{p}$ are
respectively spin and momentum operators of the valence nucleon,
$\rho(r)$ is the density of nucleons in the core normalized by the
condition $\int d\vec{r}\rho(r)=A$ (the atomic number is assumed to
be large, $A\gg 1$).

At $g_n < 10$ a simple-minded estimate for a typical weak mixing matrix element
is:
\begin{equation}\label{ca}
\langle W \rangle < 20\;eV
\end{equation}
More sophisticated calculations based on a Woods-Saxon potential with
the spin-orbit interaction produce the following upper limit on the
concrete matrix element of interest for the proposed experiment with
$^{207}$Pb:
\begin{equation}
\langle 3d\; 5/2^+|W| 2f\; 5/2^-\rangle < 14 \; eV.
\end{equation}
It is close to the expected accuracy of the experiment discussed
in Ref. \cite{sb}.  Of course, this experiment still would be both
interesting and informative, so much the more that it would be the
first occasion when PNC effects in the same nucleus were measured
both in atomic and nuclear experiments.

However, as to the hypothesis itself, according to which the value of
the weak mixing matrix element is as high as 100 eV, such a large its
value does not agree with the results of the atomic PNC experiments.

\vspace{1cm}
2. Up to now CP-violation has been observed in K-meson decays only.
One more source of the information on this phenomenon are
the upper limits on electric dipole moments (EDM) established both in
the neutron experiments and in atomic and molecular spectroscopy: due
to them a lot of models of CP-violation have been ruled out.
The best experimental upper limit on the neutron EDM $d(n)$ (the
combined result of Refs. \cite{gr,pe}) is:
\begin{equation}\label{ne}
d(n)/e < 7 \cdot 10^{-26}\,cm.
\end{equation}
Impressive results for the electron EDM were obtained in experiments with
paramagnetic atoms, cesium \cite{hunt} and thallium \cite{cd}. In
particular, the thallium experiment resulted in
\begin{equation}\label{ee}
d(e)/e = (1.8 \pm 1.2 \pm 1.0) \cdot 10^{-27}\,cm.
\end{equation}

In the standard model the neutron EDM arises to second order in $G$
only and is therefore very small. It is controlled by long-distance
contributions and constitutes \cite{kz}
\begin{equation}
d(n)/e \sim 10^{-32} - 10^{-31}\,cm.
\end{equation}
(The estimate given in Ref. \cite{gp} is an order of magnitude
larger.) Even more tiny is the electron EDM in the standard model:
\begin{equation}
d(e)/e < 10^{-40}\,cm.
\end{equation}

The highest absolute precision has been achieved in experiments with
diamagnetic atoms and molecules, mercury and thallium fluoride.  A
record-breaking upper limit on electric dipole moment of anything was
reported in \cite{kl}. The measurements of atomic EDM of the mercury
isotope $^{199}$Hg result in
\begin{equation}\label{hg}
d(^{199}Hg)/e < 9.1 \cdot 10^{-28}\,cm.
\end{equation}
Still, the upper limit on the neutron EDM following from
(\ref{hg})
\begin{equation}\label{ne1}
d(n)/e <6 \cdot 10^{-25}\,cm
\end{equation}
is an order of magnitude worse than the direct one (\ref{ne}).

However, CP-odd nuclear forces are much more effective in inducing
nuclear dipole moments than neutron or proton EDM \cite{sfk}.  Let us
present the effective CP-odd interaction of the external nucleon with
nuclear core as
\begin{equation}
W = \frac{G}{\sqrt 2}\;\frac{\xi}{2m}\;\vec{\sigma}\,\vec{\nabla}\rho(r)
\end{equation}
where $\xi$ is its dimensionless characteristic. Then the experimental
limit (\ref{hg}) corresponds to
\begin{equation}
\xi < 1.7\cdot 10^{-3}.
\end{equation}
It looks to be a serious challenge to reach a comparable accuracy in
neutron scattering experiments.

But again the standard model prediction for the constant $\xi$ does not
exceed $10^{-9}$. Taken together with the standard model predictions
for the neutron and electron EDMs, does not it mean that the
experiments discussed in this section are of no serious interest for the
elementary particle physics, are nothing else but mere exercises in
precision spectroscopy?

Just the opposite. It means that these experiments now, at the
present level of accuracy are extremely sensitive to possible new
physics beyond the standard model, physics which does not manifest
in the kaon decays.
Since various models of CP-violation have as a rule too many degrees
of freedom, it is natural to present the implications of the neutron
and atomic experiments in a purely phenomenological way: to construct
CP-odd quark-quark, quark-gluon and gluon-gluon operators of low
dimension, and find upper limits from those experiments on the
corresponding coupling constants \cite{kky}. The results are given in
Table 1

  \begin{table}[h]
  \begin{center}
  \begin{tabular}{ccc}
$k_i O_i$     &$d(n)/e<7\cdot10^{-26}\,cm$&$d(^{199}Hg)/e<9\cdot10^{-28}\,cm$\\
                     &                   &                                \\
$k_s(\bar {q_1} i\gamma_5 q_1)(\bar {q_2} q_2)$&$|k_s|<2\cdot10^{-5}$
&$|k_s|<2\cdot10^{-6}$\\
                     &                         &                          \\
$k^c_s(\bar {q_1} i\gamma_5 t^a q_1)(\bar {q_2} t^a q_2)$&   &            \\
$q_1=q_2$            & $|k^c_s|<7\cdot10^{-5}$ & $|k^c_s|<7\cdot10^{-6}$\\
$q_1\ne q_2$         & $|k^c_s|<7\cdot10^{-5}$ & $|k^c_s|<6\cdot10^{-4}$\\
                     &                   &                        \\
$k_t(1/2)\epsilon_{\mu\nu\alpha\beta}(\bar u \sigma_{\mu\nu} u)
(\bar d \sigma_{\alpha\beta} d)$&$|k_t|<8\cdot10^{-6}$&$|k_t|<7\cdot10^{-5}$\\
                     &                   &                              \\
$k^c_t(1/2)\epsilon_{\mu\nu\alpha\beta}(\bar u \sigma_{\mu\nu} t^a u)
(\bar d \sigma_{\alpha\beta} t^a d)$&$|k^c_t|<6\cdot10^{-6}$
&$|k^c_t|<5\cdot10^{-5}$\\
                     &                   &                              \\
$k^g_q\, m_p\, \bar q \gamma_5\sigma_{\mu\nu} gG^a_{\mu\nu} t^a q$
                     &$|k^g_q|<2\cdot10^{-7}$&$|k^g_q|<4\cdot10^{-8}$\\
                     &                   &                              \\
$k^g(1/6)\epsilon_{\mu\nu\alpha\beta}f^{abc}G^a_{\mu\nu}G^b_{\alpha\rho}
G^c_{\beta\rho}$     &$|k^g|<3\cdot10^{-5}$&$|k^g|<3\cdot10^{-4}$\\
                     &                   &                              \\
$\theta(\alpha_s/8\pi)(1/2)\epsilon_{\mu\nu\alpha\beta}G^a_{\mu\nu}
G^a_{\alpha\beta}$   &$|\theta|<2\cdot10^{-10}$&$|\theta|<7\cdot10^{-10}$\\
                     &                   &                         \\
                     &                   &              \\
                     & {\bf Table 1}       &                         \\
\end{tabular}
\end{center}
\end{table}
 where the limits on the dimensionless
constants $k_i$ of effective operators $$\frac{G}{\sqrt 2}\sum_i k_i
O_i$$ for the CP-odd interaction of u-, d-quarks and gluons are
presented \cite{kk}. Some upper limits following from the atomic experiment
have been derived from the bound (\ref{ne1}) extracted from the same
mercury result.

Clearly, the neutron and atomic experiments are complementary to each
other. Some effective constants are bounded by them on the
``microweak'' level or even better.

\vspace{1cm}
3. Direct experimental information on the T-odd, P-even (TOPE)
interactions is rather poor. Best limits on the relative magnitude of
the corresponding admixtures to nuclear forces lie around $10^{-3}$.
We will relate again all interactions to the Fermi weak interaction
constant $G$. Since the nuclear scale of weak interactions is
$Gm^{2}_{\pi}\sim 2 \cdot 10^{-7}$, those limits can be formulated
as $10^{4}G$. Direct experimental limits on various TOPE
interactions, together with proposals, and new limits obtained in
Refs. [26 - 28], are presented in Table 2.

  \begin{table}[h]
  \begin{center}
  \begin{tabular}{cccc}
           &   NN        &  eN         &  $\beta$-decay         \\
           &             &             &                        \\
direct limits & $<10^4G$ &             & $<0.5, \; <10^{-3}$     \\
           &             &             &                        \\
proposals  &   $<10G$    &  $<10^4G$   & $<10^{-3}$     \\
           &             &             &                        \\
new limits & $<10^{-4}G$ & $<10^{-4}G$ & $Im\,(C_S+C'_S)<4\cdot 10^{-3}$\\
           &             &             & $Im\,(C_T+C'_T)<5\cdot 10^{-4}$
\\
           &             &             & $Im\,(C_P+C'_P)<0.3$          \\
           &             &             &                              \\
           &             &{\bf Table 2}&
\end{tabular}
\end{center}
\end{table}

Let us point out first of all that the predictions of all modern
renormalizable theories of CP-violation (and not only the standard
model!) cannot exceed $(10^{-3} - 10^{-4})\,G$. The reason is obvious.
Parity violation is an intrinsic property of all these models, and
therefore T-odd, P-even effects should be roughly of the same order
of magnitude as T-odd, P-odd ones. And again, in no way does it mean
by itself that the experimental efforts in this field do not make
sense. They do, but it should be understood clearly that this is the
search for essentially new physics, well beyond the modern theories.

The approach adopted in Ref. \cite{khr} consisted in combining
phenomenological TOPE
4-fermion operators with P-odd part of the electroweak radiative
corrections. These one-loop short-distance corrections generate T- and
P-odd 4-fermion operators. Upper limits on the corresponding
effective constants are extracted from the bounds on the neutron and
atomic dipole moments. In this way one gets for different TOPE
constants the upper limits on the level
\begin{equation}\label{ls2}
(1 - 10)\; G.
\end{equation}
The estimates performed independently by M.G. Kozlov and myself
(cited in Ref. \cite{khr}) show that the account for the
long-distance effects in the interplay of the usual neutral-current
weak interaction and the discussed TOPE one leads to limits weaker
than (\ref{ls2}) obtained via the short-distance mechanism. The
result of recent elaborate papers [29 - 31] consists in fact in the
same conclusion.

Much better than (\ref{ls2}) limits presented in Table 2 were
obtained in Ref. \cite{ck} by calculating in two-loop approximation
directly electron and quark dipole moments (instead of effective T-
and P-odd 4-fermion operators)
\footnote {One cannot exclude of course the possibility that the
contributions of various particles to the two-loop diagrams discussed
cancel out. This possibility emphasized in Ref. \cite{hhm} refers
obviously to any estimates (including certainly those of Ref.
\cite{hhm} itself) made in a field where no reliable theory exists.
As to the analogy with the well-known GIM mechanism mentioned in
\cite{hhm}, it does not look relevant here.  The reasons for the GIM
cancellation in the standard model are well-known, but what have they
to do with the discussed nonrenormalizable TOPE interactions?}.

At last some information can be obtained in an analogous way
concerning even $\beta$-decay constants \cite{khr,khr1}. To relate
them to the $eN$ TOPE interaction one should evidently switch on the
W-exchange. Unfortunately, this procedure is more ambiguous than the
switching on Z-exchange used in our previous consideration.

One-loop approach leads here to the limits on the T-odd scalar,
tensor and pseudoscalar constants presented in Table 2.  The two-loop
approximation leads to the limits on the T-odd part of $\beta$-decay
interaction with derivatives on the level \cite{khr1}
\begin{equation}\label{beq}
10^{-4}\,G.
\end{equation}


\begin{thebibliography}{99}

\bibitem{sf} O.P. Sushkov, V.V. Flambaum: Pis'ma Zh.Eksp.Teor.Fiz. {\bf 32}
(1980) 377 [Sov.Phys.JETP Lett {\bf 32} (1980) 353]
\bibitem{fb} C.M. Frankle, J.D. Bowman, J.E. Bush, P.P.J. Delheij,
C.R. Gould, D.G. Haase {\it et al.}:
Phys.Rev.Lett. {\bf 67} (1991) 564; Phys.Rev.C {\bf 46} (1992) 778
\bibitem{fl} V.V. Flambaum: Phys.Rev. C {\bf 45} (1992) 437
\bibitem{bg} J.D. Bowman, G.E. Garvey, C.R. Gould, A. Hayes, M.B.
Johnson: Phys.Rev.Lett. {\bf 68} (1992) 780
\bibitem{kj} S.E. Koonin, C.W. Johnson, P. Vogel:
Phys.Rev.Lett. {\bf 69} (1992) 1163
\bibitem{ab} N. Auerbach, J.D. Bowman:
Phys.Rev. C {\bf 46} (1992) 2582
\bibitem{lw} C.H. Lewenkopf, H.A. Weidenm\"{u}ller:
Phys.Rev. C {\bf 46} (1992) 2601
\bibitem{fz} V.V. Flambaum, V.G. Zelevinsky: Preprint MSUCL-963,
December 1994; submitted to Phys.Lett. B
\bibitem{ts} L.V. Inzhechik, E.V. Mel'nikov, A.S. Khlebnikov, V.G.
Tsinoev, B.J. Ragozev: Zh.Eksp.Teor.Fiz. {\bf 93}
(1987) 800 [Sov.Phys. JETP {\bf 66} (1987) 450]; Yad.Fiz. {\bf 44}
(1986) 1370 [Sov.J.Nucl.Phys. {\bf 44} (1986) 890]
\bibitem{ts1} L.V. Inzhechik, A.S. Khlebnikov, V.G.
Tsinoev, B.J. Ragozev, M.Yu. Silin, Yu.M. Pen'kov: Zh.Eksp.Teor.Fiz.
{\bf 93} (1987) 1560 [Sov.Phys. JETP {\bf 66} (1987) 897]
\bibitem{sb} J.J. Szymansky, J.D. Bowman, M. Leuschner, B.A. Brown,
I.C. Girit: Phys.Rev. C {\bf 49} (1994) 3297
\bibitem{dkt0} V.F. Dmitriev, I.B. Khriplovich, V.B. Telitsin:
Preprint BINP 94-98, December 1994; submitted to Phys.Rev. C
\bibitem{mvm} D.M. Meekhof, P. Vetter, P.K. Majumder, S.K. Lamoreaux,
E.N. Fortson: Phys.Rev.Lett. {\bf 71} (1993) 3442
\bibitem{fk} V.V. Flambaum, I.B. Khriplovich: Zh.Eksp.Teor.Fiz. {\bf 79}
(1980) 1656 [Sov.Phys. JETP {\bf 52} (1980) 835]
\bibitem{fks} V.V. Flambaum, I.B. Khriplovich, O.P. Sushkov: Phys.Lett.
{\bf B145} (1984) 367
\bibitem{gr} K.F. Smith {\it et al.}: Phys.Lett. B {\bf 234} (1990) 191
\bibitem{pe} I.S. Altarev {\it et al.}: Phys.Lett. B {\bf 276} (1992) 242
\bibitem{hunt} L.R. Hunter: {\it in} Proceedings of the International
Conference: Neutral Currents Twenty Years Later, ed. U. Nguyen-Khac and
A.M. Lutz (World Scientific, Singapore, 1993), p. 271
\bibitem{cd} E.D. Commins, S.B. Ross, D. DeMille, B.C. Regan:
Phys.Rev. A, to be published
\bibitem{kz} I.B. Khriplovich, A.R. Zhitnitsky: Phys.Lett. B {\bf 109}
(1982) 490
\bibitem{gp} M.B. Gavela, A. Le Yaouanc, L. Oliver, O. P\`{e}ne,
J.-C. Raynal, T.N. Pham: Phys.Lett. B {\bf 109} (1982) 215
\bibitem{kl} W.M. Klipstein, S.K. Lamoreaux, B.R. Heckel, E.N.
Fortson, J.P. Jacobs: Phys.Rev. A, to be published
\bibitem{sfk} O.P. Sushkov, V.V. Flambaum, I.B. Khriplovich:
Zh.Eksp.Teor.Fiz. {\bf 87} (1984) 1521 [Sov.Phys. JETP {\bf 60}
(1984) 873]
\bibitem{kky} V.M. Khatsymovsky, I.B.Khriplovich, A.S. Yelkhovsky:
Ann.Phys. {\bf 186} (1988) 1
\bibitem{kk} V.M. Khatsymovsky, I.B.Khriplovich: Phys.Lett. B {\bf
296} (1992) 219
\bibitem{khr} I.B. Khriplovich: Nucl.Phys. B {\bf 352} (1991) 385
\bibitem{khr1} I.B. Khriplovich: Pis'ma Zh.Eksp.Teor.Fiz. {\bf 52}
(1990) 1065 [Sov.Phys.JETP Lett. {\bf 52} (1990) 461]
\bibitem{ck} R.S.Conti, I.B. Khriplovich: Phys.Rev.Lett. {\bf 68}
(1992) 3262
\bibitem{hh} W.C. Haxton, A. H\"{o}ring: Nucl.Phys. A {\bf 560}
(1993) 469
\bibitem{hhm} W.C. Haxton, A. H\"{o}ring, M. Musolf: Phys.Rev. D {\bf
50} (1994) 3422
\bibitem{egh} J. Engel, C.R. Gould, V. Hnizdo: Phys.Rev.Lett. {\bf
73} (1994) 3508

\end{thebibliography}
\end{document}